\documentclass{aa}
\usepackage{natbib,graphics}
\usepackage{times}
\bibpunct{(}{)}{;}{a}{}{,}

\newcommand{\Msol}{\mbox{$\mathrm M_{\odot}$}}
\newcommand{\Rsol}{\mbox{$\mathrm R_{\odot}$}}

\begin{document}

\sloppy
  
\title{OGLE-TR-3: A Possible New Transiting Planet\thanks{Based on  
observations collected at the European Southern Observatory, Paranal, Chile
(ESO Programme 269.C-5034).}} 
\author{S\@. Dreizler\inst{1}, P.\@H\@. Hauschildt\inst{2},
  W\@. Kley\inst{3}, T\@. Rauch\inst{1,4}, S\@.L\@. Schuh\inst{1},
  K\@. Werner\inst{1}, B\@. Wolff\inst{5}}

\offprints{S. Dreizler \\ \email{dreizler@astro.uni-tuebingen.de}}

\institute{Institut f\"ur Astronomie und Astrophysik, Abt. Astronomie,
           Sand 1, D-72076 T\"ubingen, Germany 
           \and
           Hamburger Sternwarte, Gojenbergsweg 112, D-21029 Hamburg, Germany
           \and 
           Institut f\"ur Astronomie und Astrophysik,
           Abt. Computational Physics, Auf der Morgenstelle 10, D-72076
           T\"ubingen, Germany 
           \and
           Dr.-Remeis-Sternwarte, Sternwartstra\ss e 7, D-96049 Bamberg,
           Germany 
           \and
           European Southern Observatory, Karl-Schwarzschild-Stra\ss e 2,
           D-85748 Garching, Germany 
}
\date{received; accepted}
\authorrunning{Dreizler et al.}
\titlerunning{OGLE-TR-3: A Possible New Transiting Planet}
\abstract{Recently, 59 low-luminosity object transits were reported from
  the Optical Gravitational Lensing Experiment (OGLE). Our follow-up
  low-resolution spectroscopy of 16 candidates provided two objects,
  \object{OGLE-TR-3} and \object{OGLE-TR-10}, which have companions with
  radii compatible with those of gas-giant planets. Further high-resolution
  spectroscopy revealed a very low velocity variation ($<500$\,m\,s$^{-1}$)
  of the host star \object{OGLE-TR-3} which may be caused by its unseen
  companion. An analysis of the radial velocity and light curve results in
  $M<2.5$\,$\mathrm M_{\rm Jup}$, $R<1.6$\,$\mathrm R_{\rm Jup}$, and an
  orbital separation of about 5\,\Rsol, which makes it the planet with the
  shortest period known. This allows to identify the low-luminosity
  companion of \object{OGLE-TR-3} as a possible new gas-giant
  planet. If confirmed, this makes \object{OGLE-TR-3} together with
  \object{OGLE-TR-56} the first extrasolar planets detected via their
  transit light curves.
   } \maketitle

\section{Introduction}
The detection of planets outside the solar system was a longstanding goal
of astronomy. After the first successful detections
(\citealt{wolszczan:92,mayor:95}), the search with various
methods (see \citealt{schneider:02} for an overview) was largely
intensified. Out of the 104 presently known planets, 100 have been detected
with Doppler-velocity measurements of the planets' host stars. All these
planets were found around solar-like stars. The other four are planets
around pulsars and were found by periodic pulse-modulation measurements
(see The Extrasolar Planets Encyclopaedia
http://www.obspm.fr/encycl/catalog.html for an up-to-date overview). 

Until very recently, no planet had yet been found by photometric monitoring. 
\object{HD\,209458} was the only system which has an orbital
inclination that allows the measurement of the eclipse of the host star by
the planet {\citep{charbonneau00,henry:00}. This planetary companion was,
however, known before from Doppler measurements \citep{mazeh:00}. The
current paucity of extrasolar planets found via the transit method is
deplorable for several reasons. The transiting systems provide more
reliable parameters. With the known inclination, the radial velocity
variation provides a precise mass determination and not only a lower mass
limit. Together with the radius, the density of the planet can additionally
be derived. The transit also opens, at least in principle, the possibility
to explore the planet's atmosphere during the eclipse
\citep{charbonneau02}. Furthermore, transits are the most promising way to
detect earth-like planets around solar-like stars and it is therefore the
method of choice for several ground and space based projects.

Recently, 59 transiting planet candidates were announced by the Optical
Gravitational Lensing Experiment (OGLE) consortium \citep{udalski:02, udalski:02b}. These candidates were extracted
from a sample of about 5 million stars observed during a 32-day photometric
monitoring. In a sub-sample of 52000 stars with a photometric accuracy
better than 1.5\%, these 59 candidates exhibit light curves indicating the
presence of a transiting low-luminosity companion. From this sample we
obtained low-resolution spectra for 16 objects \citep{dreizler:02} in order
to derive spectroscopic radii of the primary stars and together with the
eclipse light curve also the companion radii. While 14 low-luminosity
companions could be identified as M stars, two objects, namely
\object{OGLE-TR-3} and \object{OGLE-TR-10}, have companions with radii
compatible with those of gas-giant planets. \object{OGLE-TR-10} has been
further investigated by Hatzes et al.\, (ESO Programme 268.C-5772). We will
present in this paper our VLT follow-up observations for a dynamical mass
determination of the companion of \object{OGLE-TR-3}. 

Very recently, \cite{konacki:03} claimed the verification of the planetary
nature of \object{OGLE-TR-56}. They also investigated \object{OGLE-TR-3},
\object{OGLE-TR-10}, \object{OGLE-TR-33}, and \object{OGLE-TR-58}. While
the interpretation as planetary host star for none of these objects is
conclusive from their work yet, they 
regard \object{OGLE-TR-3} as the result of blending and grazing eclipse in
contrast to our interpretation. 

In the following, we
describe our observations and data reduction (Sect.\, 2). The analysis is
presented in Sect.\, 3 and the results are discussed in Sect.\, 4.

\begin{figure}[th]
\vspace{6.5cm}
\includegraphics{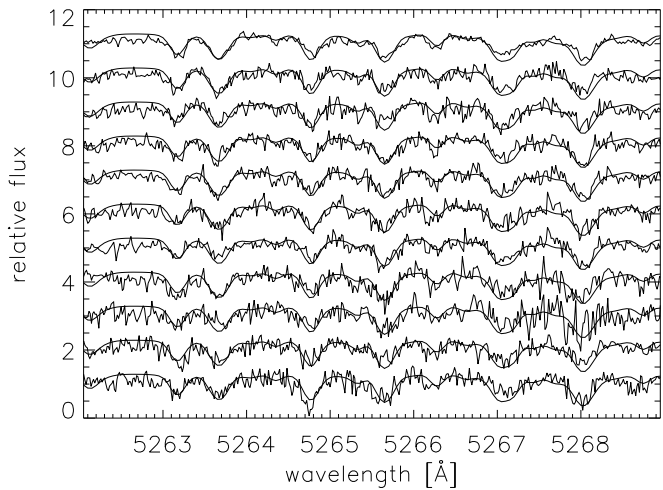}
\caption[]{From top to bottom: A small section of the co-added 
  \mbox{OGLE-TR-3} spectrum and the individual spectra of the ten observing
  blocks (1\,h observing time each) ordered by JD as listed in 
  Table\,\ref{Tobs}.}
\label{Fnormalized}
\end{figure}
\begin{figure*}[th]
\vspace{11cm}
\includegraphics{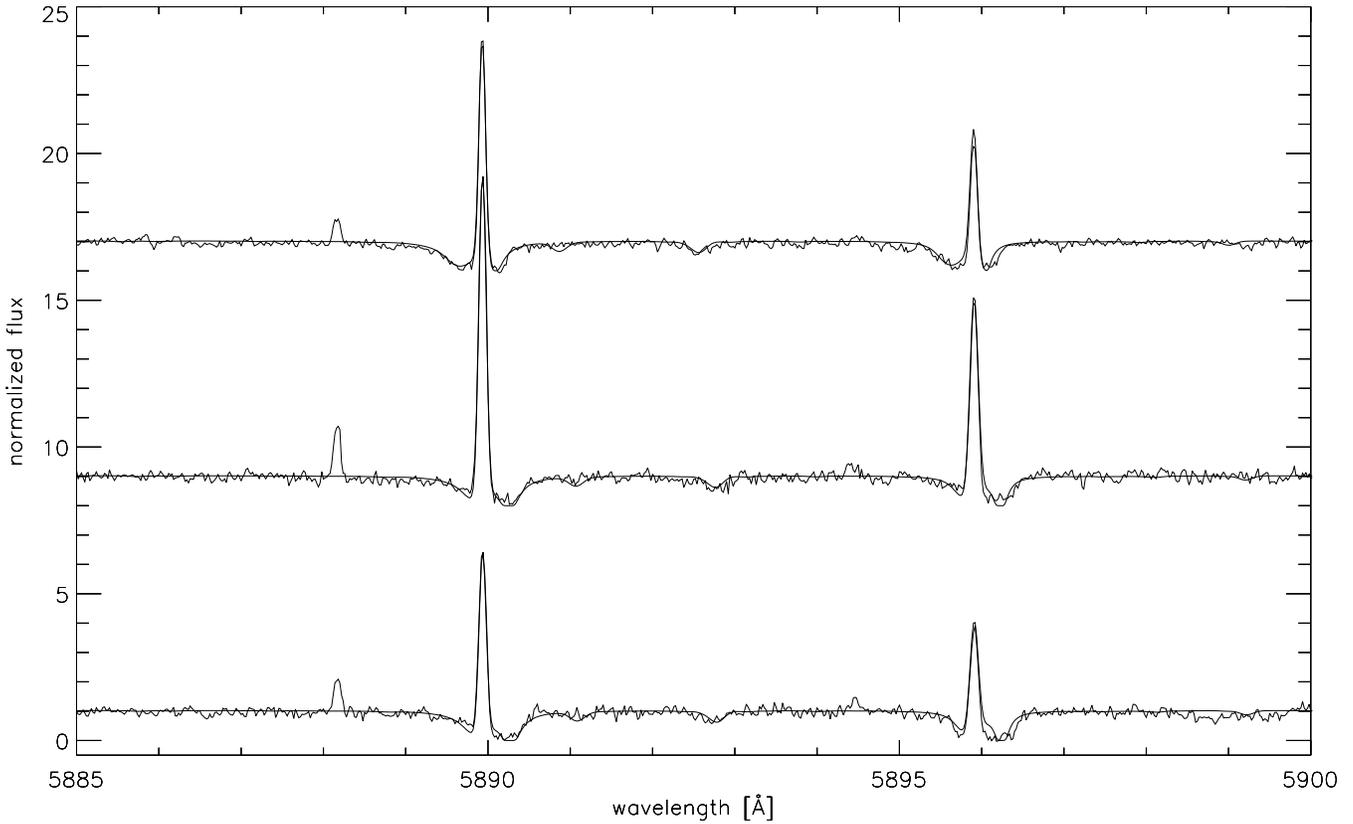}
\caption[]{Three normalized VLT-UVES spectra of OGLE-TR-3, obtained at
  JD-2452400 = 62.61590, 86.57832, and 89.56877 (from bottom to top), around the NaD
  doublet. The fit consists of three components: A synthetic
  spectrum shifted by the measured radial velocity (blue wing), a Gaussian
  emission 
  component from the earth's atmosphere, and an interstellar absorption of a
  column density of $5\cdot 10^{15}$\,cm$^{-2}$ neutral Na atoms for the
  red wing. The velocity of the interstellar component has a fixed offset
  of +20\,km\,s$^{-1}$ relative to the stellar component.
}
\label{Fspectrum}
\end{figure*}

\section{Observations and Data Reduction}
The aim of the observations is a detection of the reflex motion of
\object{OGLE-TR-3} due to the low-luminosity companion causing the observed
eclipse light curve \citep{udalski:02}. The short orbital period
($P$=1.1899 days) in combination with a spectroscopic mass of 1.1\,\Msol\
\citep{dreizler:02}, a low-mass star ($M>0.08$\,\Msol) would produce a
radial velocity variation of the order of 10\,km\,s$^{-1}$. A discrimination
between a stellar and sub-stellar object therefore needs only moderate
instrument requirements. We observed in Service Mode with UVES mounted at 
the VLT Kueyen. We used the red arm of the spectrograph with
a central wavelength of 5800\,\AA, a slit width of 0\farcs 7, and the CCD
detectors without
binning. Image slicer\,\#1 was inserted in order to obtain the full spectral
resolution of 60000 independently of the seeing.

We were granted one night of Directors Discretionary Time which we spread
in ten observing blocks of each 3$\times$20\,min exposures to minimize the
velocity smearing due to the earth's motion. We asked for five consecutive
nights with each two observing blocks around the meridian passage of
\object{OGLE-TR-3} in order to sample the orbital period equidistantly. Due
to bad weather and other targets of opportunity this schedule could not be
kept and observations were finally spread out over about one month (for
details, see Table\,\ref{Tobs}).

The spectra were extracted using procedures of the UVES context of the
ESO-MIDAS software package. The routines provide correction for bias,
flat-field, and background level. Wavelength calibration was performed with
ThAr spectra obtained immediately after each observing block. According to
the instrument handbook this provides a precision to about 100\,m\,s$^{-1}$.

From the exposure time calculator we expected a maximum S/N of about 15 in
a 20\,min exposure. The achieved S/N was significantly lower (around
10). We therefore had to co-add the three individual frames of one
observing block to each one single exposure. As an example, we show a
small section of the co-added spectrum and of all spectra from ten
observing blocks (Fig.\,\ref{Fnormalized}) overlayed with a synthetic
spectrum from our spectral analysis (see below).

\section{Analysis}
\subsection{Spectral Analysis}
An inspection of the VLT-UVES spectra of OGLE-TR-3 reveals three spectral
components. The earth's atmosphere component, consisting of the night sky
emission and telluric lines, are easy to separate since they do not follow
the radial velocity shifts due to the earth's orbital motion. Besides the
stellar absorption lines, there is a third component, mainly visible in the
NaD doublet at about 5890\,\AA, which has a constant velocity offset of
20\,km\,s$^{-1}$. The second absorption component could be caused by a stellar
companion of \object{OGLE-TR-3} or by interstellar absorption. If this were
due to a binary companion we could use the difference in radial velocity
and interpret it as orbital velocity. With the parameters of
\object{OGLE-TR-3} this would result in an orbital period of about 4
years. This component can therefore not be responsible for the observed
1.1899\,day eclipse period. If identified as binary companion we would
expect to see other spectral features at that velocity shift, if it is also
a late F to early G star as \object{OGLE-TR-3}. This is, however, not the
case. The possible stellar companion could alternatively be of later
spectral type, down to late K. It would then be about a factor of two
fainter. Since NaD is among the strongest spectral features for those
spectral types, this would explain why we cannot detect other lines. We
should, however, detect the MgH band around 5100\,\AA, which is not the
case. 

As demonstrated in Fig.\,\ref{Fspectrum}, we can fit the NaD line nicely
with an interstellar absorption caused by a column density of $5\cdot
10^{15}$\,cm$^{-2}$ 
neutral Na atoms. This is a very plausible value since
\object{OGLE-TR-3} lies about 3\,kpc in the direction of the galactic
bulge.

We also investigated the possibility of blending, first due to a second,
equally luminous star at the velocity of OGLE-TR-3. This hypothetical third
object would then be a late F to late G star, since an earlier spectral
type would dominate the spectrum and a later one would hardly contribute to
the total brightness. The eclipse could then result from a late K or early
M star orbiting one of the G stars, the shallow eclipse would just be a
fake due to the blend (G+[G+M] scenario in the following), as suggested by
\cite{konacki:03}. We therefore carefully inspected our spectra in order to
detect asymmetries in the line profiles. As demonstrated in
Fig.\,\ref{Fnormalized} with the overlayed synthetic spectrum, all
individual spectra appear symmetric and the co-added spectrum does not show
any signs of additional broadening. The effect, if present, is expected to
be directly visible since an M star companion to a G star in the short
orbital period of OGLE-TR-3 would result in a radial velocity variation of
the order of 10\,km\,s$^{-1}$ corresponding to a one pixel shift on the
detector. This would also be visible in the co-added spectrum as additional
broadening of the same order of magnitude. This is not the case (see
below). For a more quantitative investigation, we fitted 17 strong and
unblended spectral lines with Gauss profiles and calculated the difference
between the observations and the Gauss fit for the red and blue wings
separately. The total difference is shown in Fig.\,\ref{Fasymm}. Any
asymmetry would result in systematic offsets between the red and blue line
wings. Additionally we cannot find a trend with the orbital period. Our
current data set provides no hint for blending and we therefore regard the
scenario proposed by \cite{konacki:03} as unlikely in case of OGLE-TR-3.

The second blending scenario is a close M+M star binary orbiting at
2$\times$1.1899\,days blended by a G star with a large enough distance
towards the M star binary so that it has an undetectable orbital velocity
(G+[M+M] scenario in the following). We inspected our UVES spectra in order
to detect spectral features from a possible M star. There is no evidence
for it but since the contribution of two M stars is only of the order of
3\% in the red, we cannot rule out this completely. The light curve from
such a system is discussed below (Sect. 3.3).

The third blending scenario is a line of sight blending: A G0 star is
contributing most of the light while an eclipsing binary system is situated
at a much larger distance (G+[X+Y] in the following). Its contribution to
the spectrum must be (very conservatively) less than about 30\% because we
would detect it otherwise, as discussed above. Its influence on the light
curve is also discussed below (Sect. 3.3).

\begin{figure}[th]
\vspace{6.5cm}
\includegraphics{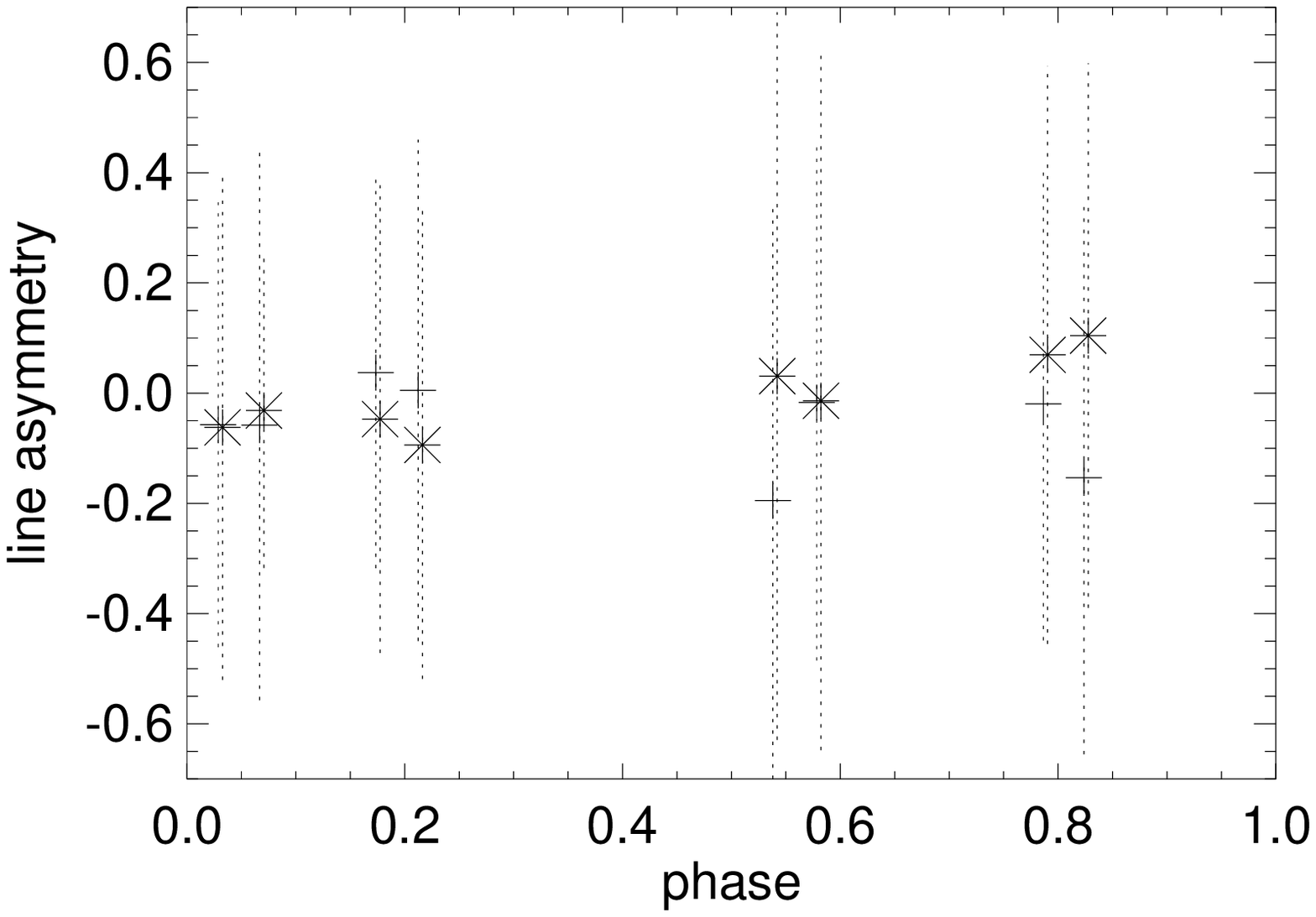}
\caption[]{Co-added differences between red ``$+$'' and blue ``$*$''
  wings of observed spectral lines and Gauss profiles with 1\,$\sigma$ error
  bars. The spectra from JD=86.53247 and 86.57832 are omitted due to very large
  uncertainties. No indication of line asymmetry is present in our data set.}
\label{Fasymm}
\end{figure}
\begin{figure*}[th]
\vspace{12.9cm}
\includegraphics{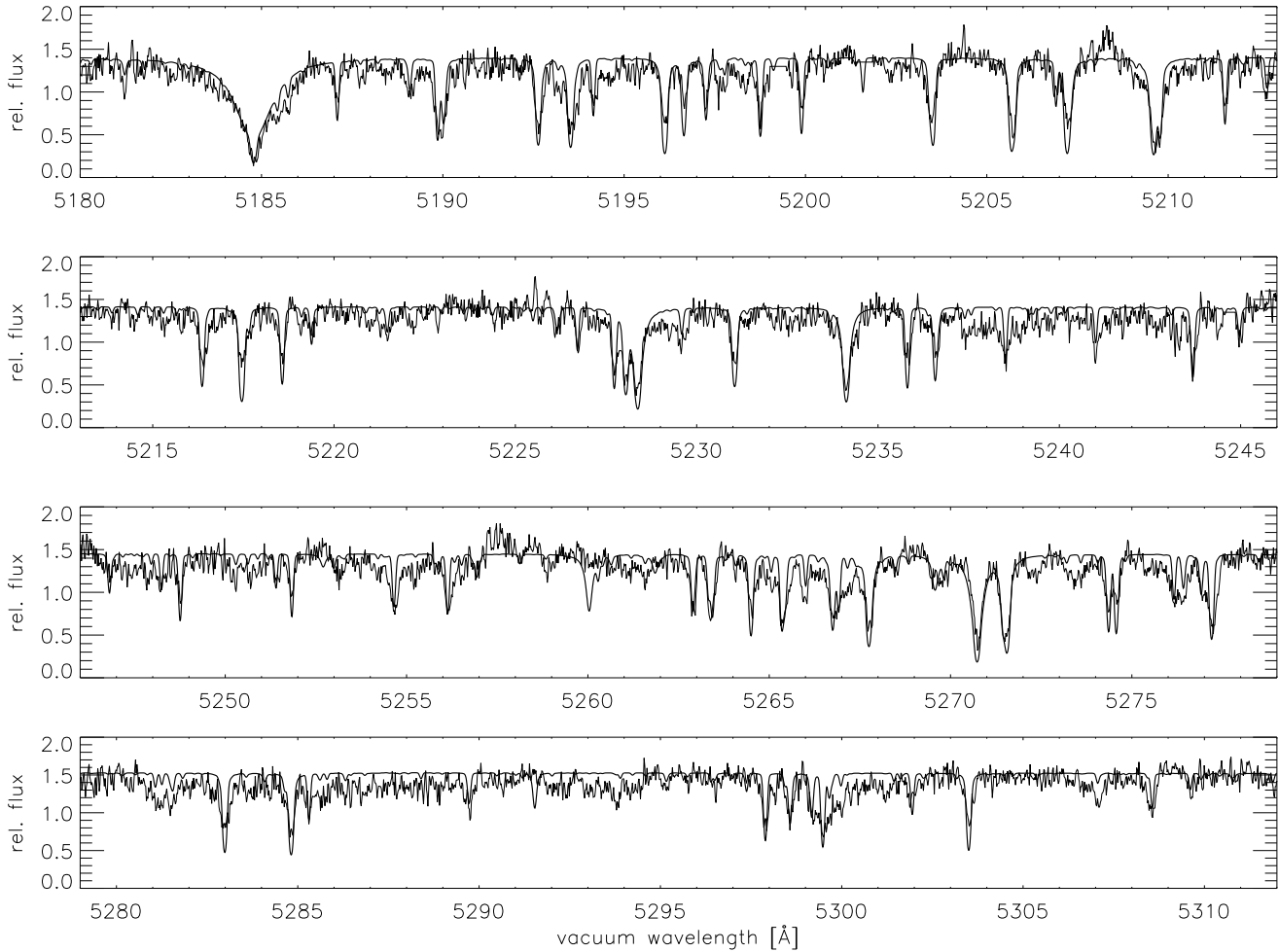}
\caption[]{Four 50\,\AA\ sections of the co-added spectrum of OGLE-TR-3
  compared to a theoretical model: $T_{\rm eff}=6100\,$K, $\log g =4.5$\,
  (cgs), and sub-solar abundances ($[{\rm M/H}]=-0.5$).} 
\label{Fspecfit}
\end{figure*}
\begin{figure*}[th]
\vspace{10.4cm}
\includegraphics{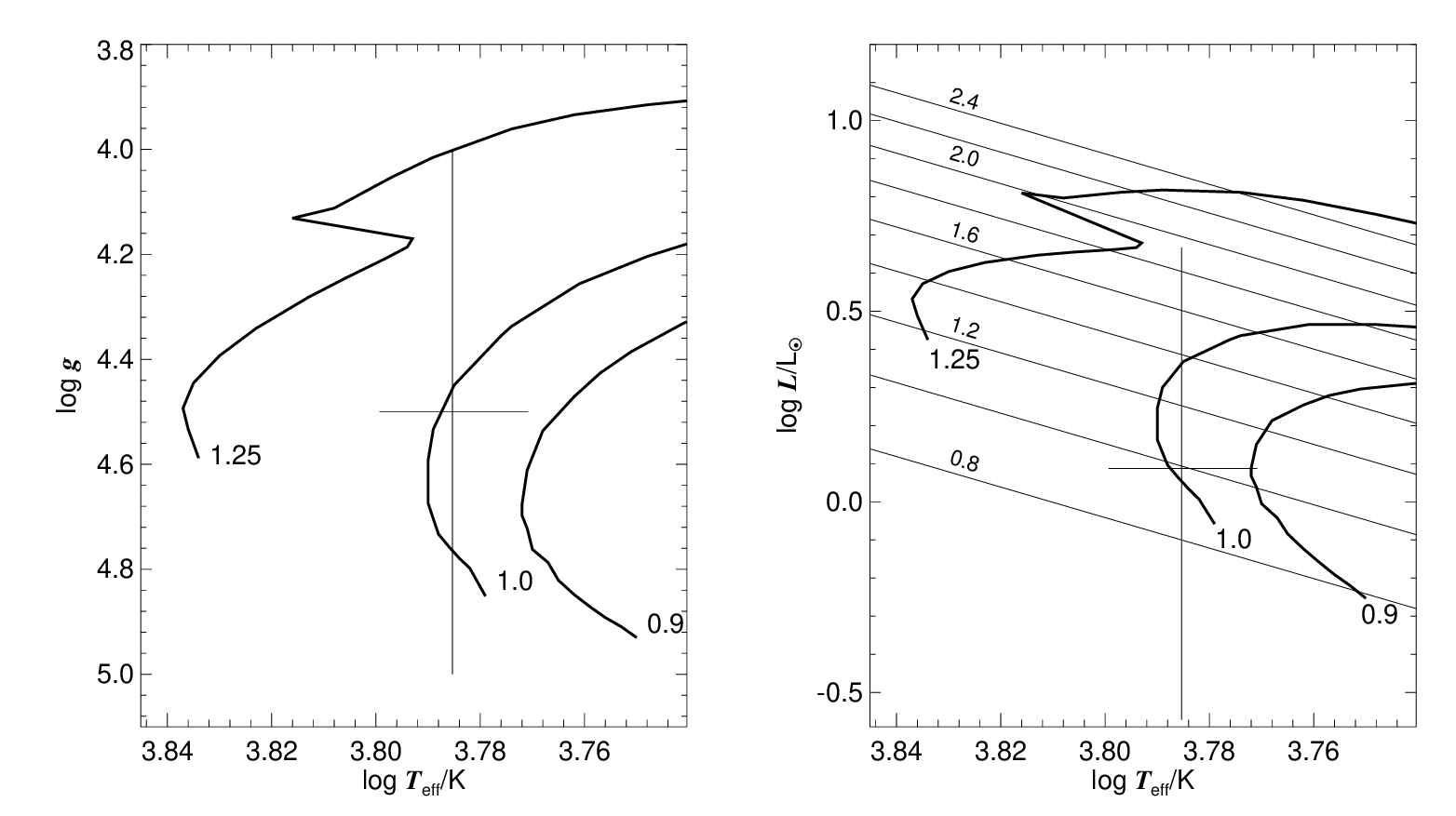}
\caption[]{The parameters of OGLE-TR-3 compared to stellar evolution models
  \citep{schaerer:93}. The evolution tracks as well as the lines of
  constant radii are labeled in solar units.}
\label{Fhrd}
\end{figure*}

We fit the co-added spectrum of OGLE-TR-3 by comparing it to a {\tt
PHOENIX} (version 13.00) spherical model atmosphere grid and synthetic
spectra calculated at the resolution of the observed spectrum
(Figs.\,\ref{Fnormalized} and \ref{Fspecfit}).  The model atmosphere grid covers a wide range of
effective temperatures, gravities, and abundances.  It includes the latest
updates with respect to the NextGen \citep{ng-giants} and Dusty/Cond
\citep{LimDust} grids. For simplicity, we have used LTE models, since NLTE
effects in the relevant parameter range can be expected to be small and
were thus neglected for the initial fits. The updates include improvements
in the equation of state data used and better molecular line lists, however,
the latter are not important for OGLE-TR-3.

The synthetic spectra grid covers a range of $6000\,$K$ \le T_{\rm eff} \le
6200\,$K and $4.0 \le \log g \le 5.0$ for solar abundances and $[{\rm
M/H}]=-0.5$, in agreement with our earlier spectral classification as
F9/G0\,V star \citep{dreizler:02}.  Synthetic spectra were computed with
$0.01\,$\AA\ steps and rotationally broadened for $v_{\rm rot}\sin i
=2\,$km\,s$^{-1}$. Again, this is a clear indication that no third object
as in the G+[G+M] scenario is present, as discussed above. In addition, the
spectral resolution was downgraded to $0.02\,$\AA\ before comparison with
the data. The observed spectra are not flux calibrated, so the synthetic
spectra were normalized to the continuum. With this procedure, we find that
the best fit parameters for OGLE-TR-3 are $T_{\rm eff}=6100\pm 200\,$K,
$\log g =4.5\pm 0.5$ and abundances between solar and $[{\rm
M/H}]=-0.5$. We did not attempt to improve the fit by changing the
abundance pattern, but inspection of the comparisons shows that this could
lead to improved fits. Models outside the parameter range do not lead to
improved fits, and were not considered further.  In Fig.\,\ref{Fhrd} we
compare these parameters with stellar evolution models \citep{schaerer:93}
and derive the stellar mass and radius.  The spectral analysis corroborated
the stellar parameters of the primary derived earlier ($R=1.05\,$\Rsol,
$M=1.1\,$\Msol\ \citealt{dreizler:02}).  However, our fit favors slightly
lower parameters, $R=1.0^{+0.9}_{-0.2}\,$\Rsol\ and
$M=1.0^{+0.3}_{-0.1}\,$\Msol. We adopt $R=1.0\,$\Rsol\ and $M=1.0\,$\Msol\
in the following radial velocity and light curve analysis.

\subsection{Radial velocity variation}
\begin{table}[ht]
\caption{Observing dates, orbital phase of OGLE-TR-3, measured radial
  velocity relative to a co-added spectrum, earth's orbital and earth's
  rotational velocity components towards OGLE-TR-3, radial velocity of
  OGLE-TR-3, and standard deviation of the measurements. The radial velocity
  column provides the measured velocity corrected for earth's motions minus
  the median of the 10 values.}
\begin{tabular}{cc|rrrrc}
\hline
%JD-2452400 & phase & RV spec. & RV earth orbit & RV earth rot.& RV OGLE-TR-3 & $\sigma$\\
%           &       & [km\,s$^{-1}$]           & [km\,s$^{-1}$]         & [km\,s$^{-1}$]       & [km\,s$^{-1}$]        & [km\,s$^{-1}$] \\
JD-2452400 & \hspace{-3mm}phase & \multicolumn{5}{c}{radial velocity [km\,s$^{-1}$]}\\
           &       & spec. & orbit & rot. & TR-3 & $\sigma$ \\
\hline
62.61590 & \hspace{-3mm}0.1754 &  0.06 &  -8.32 &  0.11 &  0.151 & 0.12\\
62.66215 & \hspace{-3mm}0.2143 &  0.15 &  -8.34 &  0.01 &  0.116 & 0.17\\
63.63365 & \hspace{-3mm}0.0307 &  0.51 &  -8.80 &  0.07 &  0.072 & 0.10\\
63.67882 & \hspace{-3mm}0.0687 &  0.56 &  -8.82 & -0.03 & -0.003 & 0.16\\
66.61967 & \hspace{-3mm}0.5400 &  2.11 & -10.20 &  0.08 &  0.290 & 0.25\\
66.66758 & \hspace{-3mm}0.5804 &  1.95 & -10.22 & -0.02 & -0.004 & 0.13\\
86.53247 & \hspace{-3mm}0.2741 & 10.11 & -18.65 &  0.15 & -0.094 & 0.58\\
86.57832 & \hspace{-3mm}0.3127 & 10.32 & -18.67 &  0.05 &  0.000 & 0.64\\
89.52445 & \hspace{-3mm}0.7885 & 11.14 & -19.77 &  0.15 & -0.192 & 0.44\\
89.56877 & \hspace{-3mm}0.8257 & 11.13 & -19.79 &  0.06 & -0.306 & 0.55\\
\hline
\label{Tobs}
\end{tabular}
\end{table}

For the determination of the radial velocity variation of
\object{OGLE-TR-3} as reflex motion due to the unseen companion, we used the
first spectrum as reference and determined the cross-correlation in
velocity space relative to the other nine. We identified twelve 50\,\AA\
intervals within the available spectral range (4800-6800\,\AA) with little
contamination from telluric lines. Additionally, we had to avoid the
spectral region between 5750 and 5850\,\AA\ where the spectrum falls onto
the gap between the two CCDs. The cross-correlation function is
approximated by a Gauss profile to determine the velocity shift at
sub-pixel precision (see Fig.\,\ref{Fccorr}). The median of the twelve
segments is then taken as the velocity shift of the spectrum. The standard
deviation is used as quality attribute of the measurement and directly
reflects the S/N of the spectra. Afterwards we co-added all spectra in
velocity space and repeated the procedure, now with the co-added spectrum
as reference. Reassuringly, neither the cross-correlation with the co-added
spectrum nor with a model spectrum does change the results significantly.

\begin{figure}[th]
\vspace{6.5cm}
\includegraphics{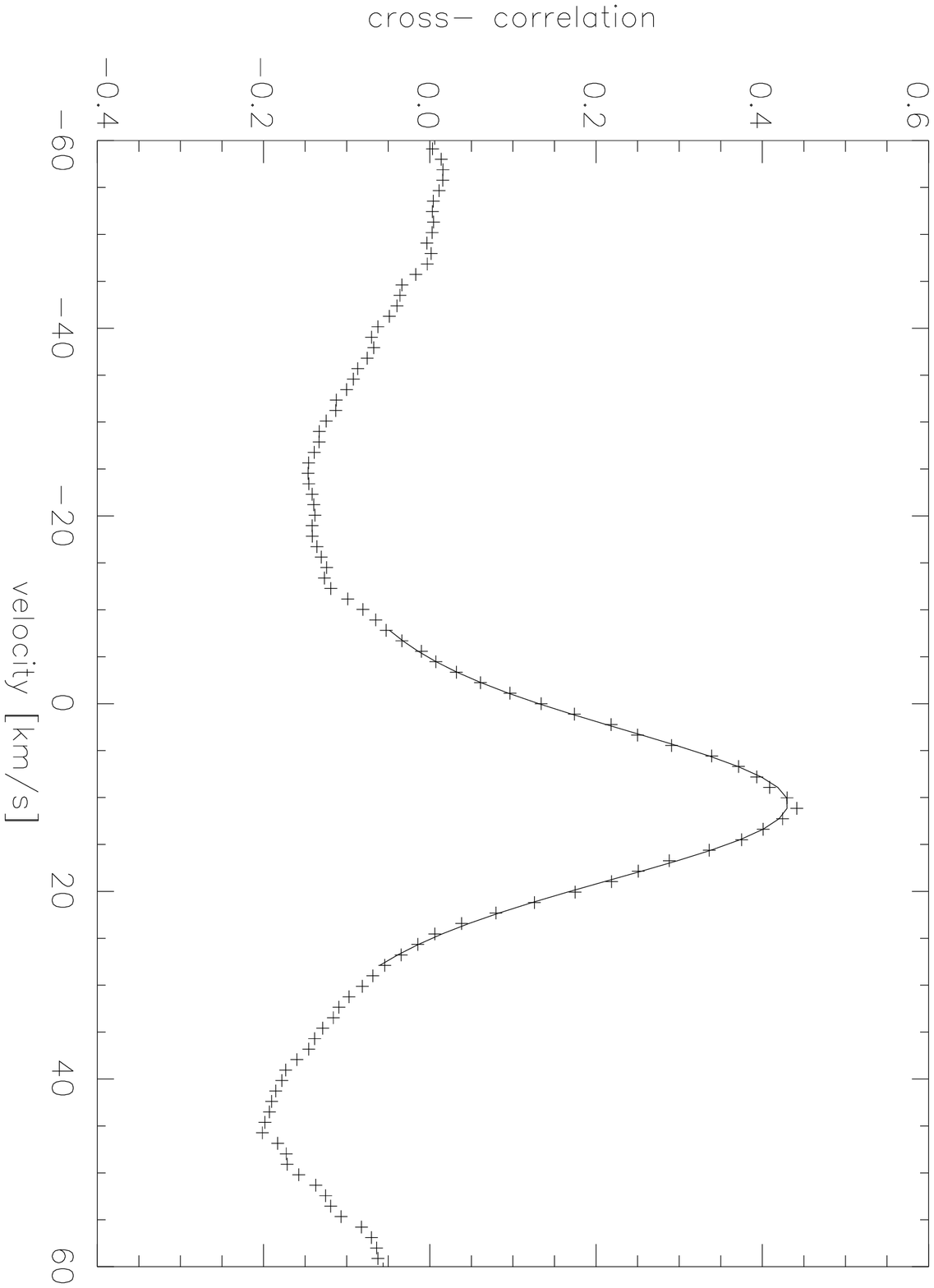}
\caption[]{An example of the cross-correlation. Co-added spectrum versus 
  the one from JD=2452486.53247 in the interval [5350\,\AA - 5400\,\AA]
  (++++) and a Gauss fit (solid line).
}
\label{Fccorr}
\end{figure}

Using the period ($P$=1.1899\,days) and epoch of the mid eclipse
(JD=2452060.22765) from \cite{udalski:02}, we calculated the phasing of each
mid of exposure\footnote{The slightly different values given at the
    OGLE-web site http://sirius.astrouw.edu.pl/$\sim$ogle result in the
    same conclusions.}. For each observation we also calculated the velocity
components of the earth towards \object{OGLE-TR-3} using the
\texttt{BARYVEL} routine available in the Interactive Data Language (IDL)
\texttt{astrolib} from Goddard Space Flight Center. It should
be noted, that this routine provides the velocity in a right-handed
coordinate system with the +X axis toward the vernal equinox, and +Z axis
toward the celestial pole. A positive velocity component towards an object
therefore produces a blue shift whereas the spectroscopic definition for
the sign of the velocity is the opposite. We also calculated the velocity
components due to the earth's rotation, which is, however, of minor
importance. In Table\,\ref{Tobs} we summarize our results for each
exposure.

An independent period determination is impossible from our few scattered
measurements. Hence we mapped the radial velocity curve to the orbital
phase (Fig.\,\ref{Frv}) using the period and zero point from
\cite{udalski:02}. These data were fitted with a sine curve with an offset, 
the velocity amplitude, and a phase shift as free parameters.  We can use
the fitted phase as an indication that the period is present in our
data. As expected, the derived radial velocity is indeed zero at mid eclipse.

\begin{figure}[th]
\vspace{6.2cm}
\includegraphics{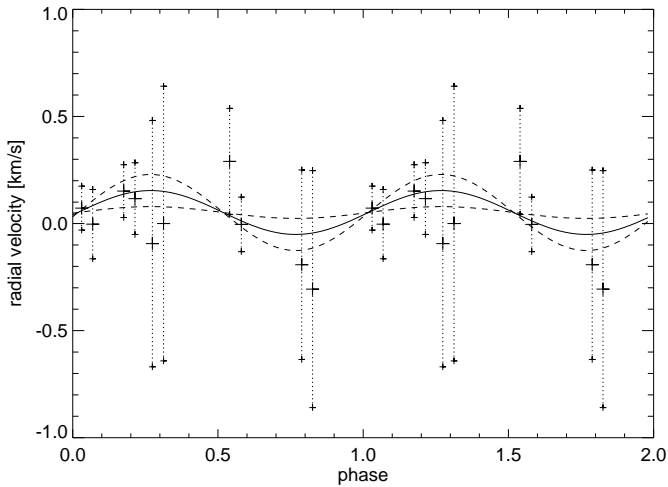}
\caption[]{Measurements of the radial velocity of OGLE-TR-3. Error bars
  indicate 1\,$\sigma$ deviations. The solid line is the best fit from our
  $\chi^2$-minimization, the dashed lines are the 3\,$\sigma$ deviations of the
  amplitude.
}
\label{Frv}
\end{figure}

\begin{figure}[th]
\vspace{6.2cm}
\includegraphics{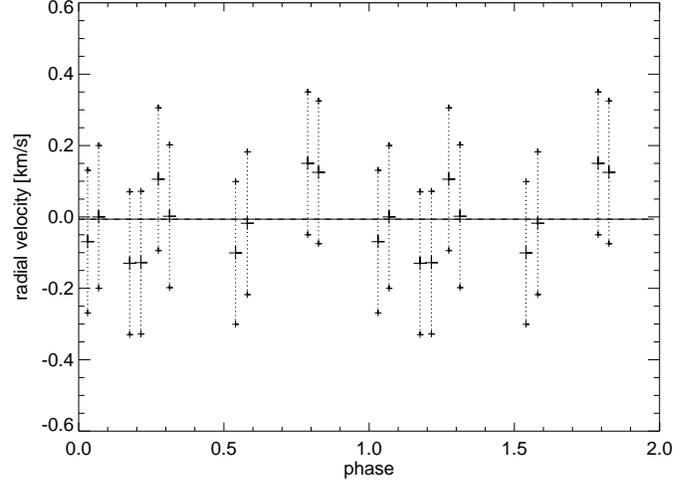}
\caption[]{Measurements of the velocity shift of the night sky emission  
  lines in the spectra of OGLE-TR-3. Error bars
  indicate 1\,$\sigma$ deviations. The solid line is the best fit from our
  $\chi^2$-minimization. 
}
\label{Fsky}
\end{figure}

First, we used an inversion based $\chi^2$-minimization
routine. Fig.\,\ref{Frv} shows the measured radial velocities with error
bars of 1\,$\sigma$ overlayed with the best fit (solid). We also show the
velocity curve for 3\,$\sigma$ variation (dashed) in the amplitude (see also
Table\,\ref{Tresults}). The resulting amplitude of about 100\,m\,s$^{-1}$ is
smaller than the median 1\,$\sigma$ error, we therefore investigate the
significance of the result in more detail.

\begin{table}[ht]
\caption{Results of the radial velocity curve fits. Left block: Results
  from the inversion based $\chi^2$-minimization. Right block: Results
  from the genetic algorithm optimization. The blocks give the best fit
  value and lower and higher limits. Note that the 3\,$\sigma$ deviation of
  the \texttt{pikaia} fits are smaller than 1\%.}
\begin{tabular}{lccc|ccc}
\hline
parameter\rule{0cm}{4mm} & \multicolumn{3}{c|}{$\chi^2$} &
\multicolumn{1}{c}{\texttt{pikaia}} \\ 
          & best & -3 $\sigma$ & +3 $\sigma$ & best \\%& -3 $\sigma$ & +3 $\sigma$\\ 
\hline
amplitude [km\,s$^{-1}$] & 0.10 & 0.08 & 0.13 & 0.12 \\%& 0.12 & 0.12 \\
phase               & 0.02 & 0.99 & 0.04 & 0.94 \\%& 0.94 & 0.94 \\
offset      [km\,s$^{-1}$] & 0.05 & 0.04 & 0.06 & 0.05 \\%& 0.05 & 0.05 \\
\hline
\label{Tresults}
\end{tabular}
\end{table}

We used night-sky emission lines which do not show velocity variation due
to the earth's motion. The resulting ``velocity'' variation
(Fig.\,\ref{Fsky}) reveals a fit with vanishing amplitude, as expected. It
also demonstrates that we recover the velocity precision of UVES ($\approx
100$\,m\,s$^{-1}$). This test is an indication that the derived velocity
amplitude is indeed due to intrinsic variations rather than to measurement
errors.

\begin{figure*}[th]
\vspace{10.4cm}
\includegraphics{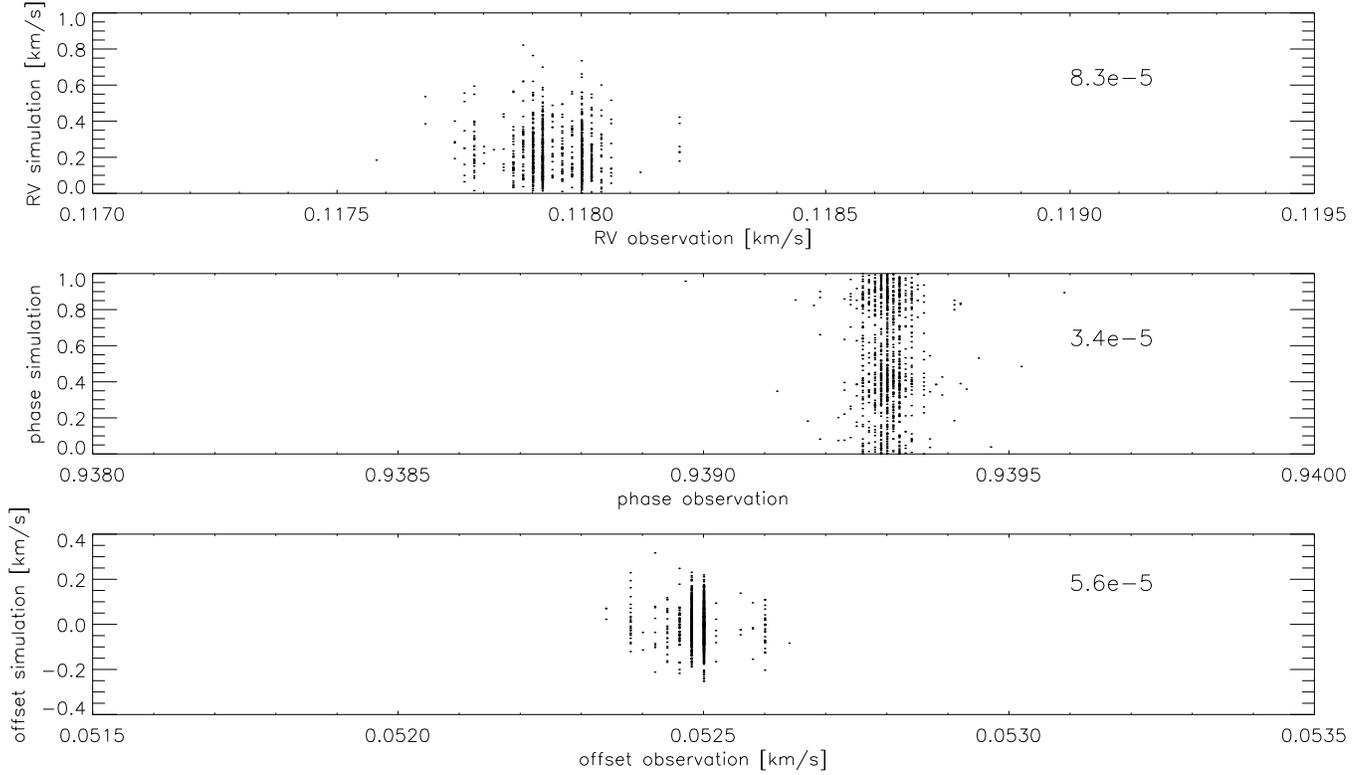}
\caption[]{Fit results of synthetic white noise curves versus fit  
  results of the OGLE-TR-3 radial velocity (RV) curve using
  \texttt{pikaia}. While the synthetic data (ordinate) spread over a wide
  interval, the real data (abscissa) are confined to a narrow range
  only. The ratio between the distribution widths of the fits is indicated
  in the plots.}
\label{Fpikaia}
\end{figure*}

Additionally, we utilized the genetic algorithm \texttt{pikaia}
\citep{charbonneau:95} to confirm the three fit parameters. The genetic
algorithm starts from randomly distributed parameter sets and increases the
quality of the fit according to selection criteria adopted from the biological
process of evolution. It therefore provides the global $\chi^2$-minimum
much more reliably than classical $\chi^2$-minimization. In order to obtain
a measurement for the reliability of the result, we ran \texttt{pikaia} with
1000 different initializations of the starting population. The deviation of
the final fit parameters is very small (Table\,\ref{Tresults}) and they are
consistent with the one of the classical $\chi^2$-minimization.

As a third test, we synthesized 1000 white noise curves with the variance
of that from the radial velocity curve and with identical timing and
fitting weights. These data sets were fitted with \texttt{pikaia}. In
Fig.\,\ref{Fpikaia} we compare the results of the \texttt{pikaia} fit of
the radial velocity curve with those from the simulations. While the
parameters of the radial velocity variations are recovered with a very
small scatter, the noise simulations produce fit results over a large
parameter space. While the fitted amplitudes and offsets are typically of
the order of the mean standard deviation of our measurements ($\approx
200$\,m\,s$^{-1}$), the phase fits are equally distributed over the interval
[0,1]. As measurement for the reliability of the fit, we provide the widths
of the parameter interval for the radial velocity fits divided by the width
of the simulation fits.

Concluding, we can fit the observed radial velocity curve with an amplitude
of about 120\,m\,s$^{-1}$. We provided several indications that this fit
indeed represents a true reflex motion of the late F/ early G type
planetary host star due to the unseen companion. This contradicts the
G+[M+M] scenario, where the G star would have a very small radial velocity
variation at a much larger orbital period.

%Even if this fit turns out to be wrong, the upper limit given
%by the standard deviation of our measurements is small enough (500\,m\,s$^{-1}$) to
%draw reliable conclusions.

\subsection{The light curve revisited}
\begin{figure}[th]
\vspace{6.5cm}
\includegraphics{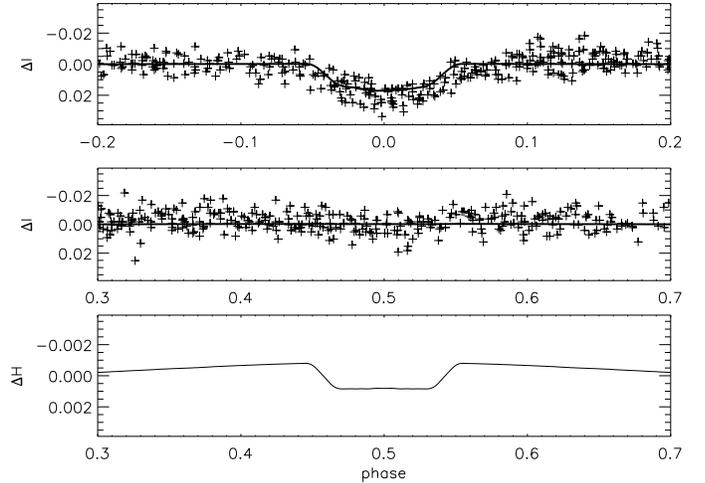}
\caption[]{Light curve of OGLE-TR-3 from \cite{udalski:02}: Top and middle:
  I band observed (++++) 
  and synthetic light curve (solid line). Bottom: Predicted H band light
  curve at a smaller scale. The primary eclipse is shown in the top panel,
  the secondary eclipse in the lower two. 
}
\label{Flc}
\end{figure}

For a consistency check, we fitted the light curve of \object{OGLE-TR-3}
published by \cite{udalski:02} with our derived parameters ($R=1.0\,$\Rsol,
$M=1.0\,$\Msol). If we estimate the temperature of the secondary due to
irradiation, we derive about 2100\,K for a 6100\,K primary and 5\,\Rsol\
separation. We utilized the binary eclipse simulation program
\texttt{nightfall} (R. Wichmann, Landessternware Heidelberg, Germany) which
calculates synthetic light curves taking into account the distortion of the
stars in Roche geometry. As can be seen in Fig.\,\ref{Flc}, we can nicely
reproduce the observed I-band light curve with an inclination of
$90^{\circ}$, a primary radius of $R=1.25\,$\Rsol\ and a secondary radius
of 0.14\,\Rsol\ (80\% filling factor of the Roche potential of a
0.0006\,\Msol\ planet). The radial velocity of the primary is
131\,km\,s$^{-1}$ for such a system, fully consistent with our determined
value. The rotational velocity of the primary is 2\,km\,s$^{-1}$ which is
far below synchronization ($\approx$5\%). Due to the scatter in the
photometry a simultaneous fit of the inclination and the secondary radius
is not very well constrained. An inclination of $85^{\circ}$ requires a
secondary radius of 0.16\,\Rsol, $80^{\circ}$ require 0.17\,\Rsol. The
formal fit quality, however, favors an inclination very close to
$90^{\circ}$. While we derive a radius ratio consistent with that from
\cite{udalski:02}, our absolute radii are slightly smaller than their
values ($R_1$=1.48\,\Rsol, $R_2$=0.18\,\Rsol), but within our
uncertainty. A very small secondary eclipse might be present in the current
photometry. Whether this is real remains to be checked with much more
precise photometry. Fig.\,\ref{Flc} also shows the predicted H-band light
curve of the secondary eclipse which is of the order of 1\,mmag. It should
therefore be possible to detect the planet directly with high precision IR
photometry. We also investigated the effect of the distortion of the
  primary due to the gravitational field of the secondary. This effect
  becomes visible in a $\sin(2P)$-variation outside the eclipses. We fitted
  both our model solution as well as the observation and find a 0.15\,mmag
  and 0.3$\pm0.3$\,mmag variation (3 $\sigma$ error), respectively. Our
  solution is therefore consistent with the observations.

\begin{figure}[th]
\vspace{4.5cm}
\includegraphics{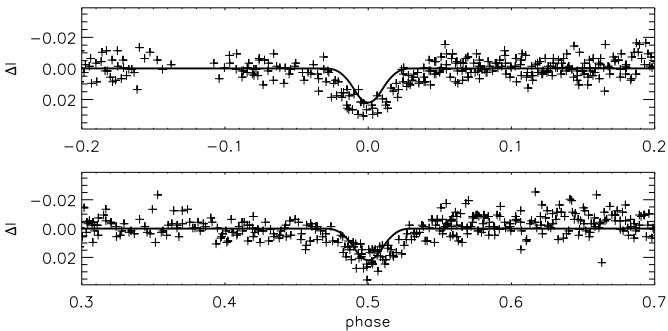}
\caption[]{Light curve of OGLE-TR-3 from \cite{udalski:02}: I band observed
  (++++) and synthetic light curve (solid line). The photometric data are
  folded with 2$\times$1.1899\,days in order to simulate the G+[M+M]
  scenario. The eclipsing system (primary minimum top, secondary minimum
  bottom panel) consists of two M2 stars; the third light contribution is that
  from a G star.
}
\label{Flc2}
\end{figure}
While our radial velocity measurements are in favor of the interpretation
of OGLE-TR-3\,B as planet, the light curve might be explained in the
G+[M+M] scenario. The eclipse of the two M2 stars diluted by the G0 star
contribution would result in a shallow eclipse of the observed magnitude
difference. It requires that the two M stars are nearly identical. The
measured period would correspond to half an orbital phase and the true
period of the system would then be 2$\times$1.1899\,days. A period search
in the original data, however, results in the largest peak at the period of
1.1899\,days. If we simulate the light curve of this system including the
third light contribution, we obtain a reasonable fit (Fig.\,\ref{Flc2}),
which is, however, not as good as in the case of a transiting planet
(Fig.\,\ref{Flc}). The eclipses are too narrow, the formal fit error is
therefore larger than for the planet transit light curve.  An increase of
the stellar radii to increase the eclipse widths would also influence the
depths since the binary would contribute more to the total light. We
therefore regard this interpretation for the OGLE-TR-3 system as
unlikely. The G+[M+M] scenario can easily be checked, since the M star
contribution to the total light is strongly wavelength dependent. Hence, we
would expect a strong color effect in the light curve.

The G+[X+Y] blending scenario would also require a false detection of the
radial-velocity variation. Furthermore, we checked possible blends for
their consistency with the light curve. Assuming that the secondary is much
fainter than the primary (no detectable secondary eclipse), we can pick
a spectral type of the secondary with the corresponding radius and mass for
a main sequence star. Unblended light curves for various contributions from
a less distant G0 star are reconstructed, the depth of the primary eclipse
then provides the radius 
of the primary. The corresponding mass together with the period is used to
derive the orbital separation and velocity (Kepler's third law, assuming
circular orbits). The time between first and fourth contact can then be
obtained from the orbital velocity and the sum of the radii, the time
between second and third contact from the orbital velocity and the
difference of the radii. In Fig.\,\ref{Fblend} these eclipse times are
plotted in units of the corresponding times obtained from the observed
light curve. A consistent solution is only possible in a very limited
parameter range: The foreground G0 star provides about 80\% of the total
light, the eclipsing binary system is then a F9 secondary and a B9
primary. However, this scenario can be ruled out when a theoretical light
curve is calculated with these parameters. The close vicinity of the two
stars would result in an observable sinusoidal variation. Additionally, the
light curve can only be reproduced very roughly with this solution. We
therefore reject the G+[X+Y] scenario based on our current observational data.
\begin{figure}[th]
\vspace{6cm}
\includegraphics{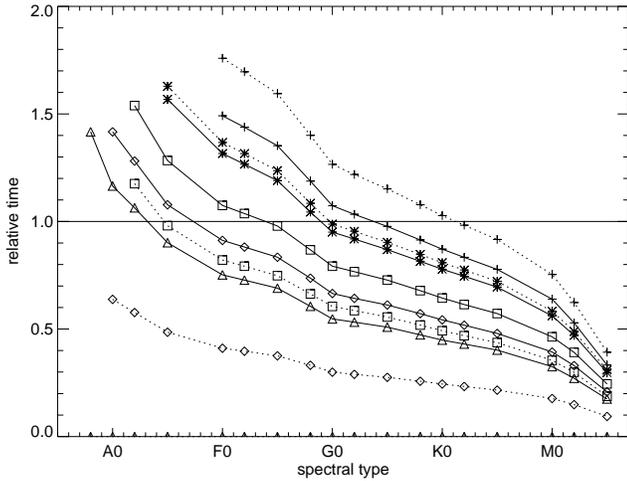}
\caption[]{Predicted eclipse durations for blending scenarios as function 
  of the spectral type of the secondary (first to fourth contact: full line;
  second to third contact dashed line) in units of eclipse durations of
  OGLE-TR-3. Contribution from a foreground star with 70, 80, 90, 95, 98\%
  is labeled with crosses, stars, squares, diamonds, and triangles
  respectively.
}
\label{Fblend}
\end{figure}

\section{Summary and Discussion}
Summarizing, our data favor the interpretation of OGLE-TR-3 as planetary
host star and we have detected the radial velocity variation due to the
reflex motion of the unseen planetary companion.

From Kepler's third law together with momentum conservation the companion
mass can be determined from the period (1.1899\,days,
\citealt{udalski:02}), the mass of the primary ($1.0\,$\Msol, this paper),
the velocity amplitude ($120$\,m\,s$^{-1}$, this paper), and the
inclination, which must be close to $90^{\circ}$ (see above, in the
following we will use $90^{\circ}$).

The upper limit of the derived radius ($R=0.17\,$\Rsol) of the unseen
companion would be consistent with an M\,star down to a gas-giant
planet. With our VLT observations, however, we can clearly rule out an
M\,star since a 0.08\,\Msol\ star would cause a 15\,km\,s$^{-1}$ reflex
velocity variation. We can also rule out a brown dwarf companion since a
brown dwarf at the lower limit for deuterium burning (0.013\,\Msol) would
still result in 2.5\,km\,s$^{-1}$ velocity amplitude.  As a conservative
upper limit for velocity variations (3\,$\sigma$) we derive
500\,m\,s$^{-1}$ which translates into 0.0026\,\Msol. If our fit is
regarded as reliable, we obtain a mass of 0.0005 to 0.0006\,\Msol\ or about
half a Jupiter mass. Due to the small period, the separation of the two
objects is only about 5\,\Rsol\ which is however still a factor of two
above the Roche stability limit. The companion's mean mass density is
comparable to the one of \object{HD\,209458}\,B. If we use our
0.0005\,\Msol\ and 0.014\,\Rsol, it amounts to 0.25\,g\,cm$^{-3}$, which is
about 20\% of the mean solar density.

The shortest known orbital period of all extrasolar planets and the very
close orbit of OGLE-TR-3 raises the question of the influence of tidal
effects between the star and planet. As the star appears to be rotating
much slower than the planetary orbital period, any tides raised by the
planet on the star will lead to an additional shrinking of the orbit.
Following the argmuments of \cite{goldreich:66} we
estimate the timescale $a/\dot{a}$ for a change in the semi-major axis of
the planet, using the present orbital parameter.  Using a value of $10^7$
for the dissipation factor $Q_*$ of the star we obtain a damping timescale
of about $2 \cdot 10^9$yrs.  As the $Q$ value for stars is quite uncertain
and may easily be higher, we may conclude that a planet is just about able
to survive in such a close orbit for the lifetime of the star.

We therefore conclude that we confirmed the low-luminosity companion of
\object{OGLE-TR-3} as an extrasolar planet ($M_{1}$=1.0\,\Msol,
$M_{2}$=0.0005\,\Msol\ and $R_{2}$=0.14\,\Rsol) with quite similar
properties compared to \object{HD\,209458} ($M_{1}$=1.05\,\Msol,
$M_{2}$=0.00066\,\Msol, $R_{2}$=0.147\,\Rsol) and \object{OGLE-TR-56}
($M_{1}$=1.04\,\Msol, $M_{2}$=0.00086\,\Msol, $R_{2}$=0.13\,\Rsol)
regarding the parameters of the primary star and planetary companion. The
system of \object{OGLE-TR-3} is slightly more extreme regarding the
separation.

Finally, if confirmed, our results for \object{OGLE-TR-3}, together
with those for \object{OGLE-TR-56}, make these objects the first two
extrasolar planets detected via the transiting method. A further
confirmation of the planetary nature would benefit from several supporting
observations: Optical B,V,R photometry to exclude a color effect expected
in the blending scenarios, IR spectroscopy to exclude an M star contribution,
H band photometry to detect the predicted very shallow secondary eclipse,
and finally, more radial velocity measurements to detect the orbital period
in the radial velocity data directly.

\begin{acknowledgements}
We want to express our thanks to the OGLE team lead by A. Udalski, who
provided their photometry which is the basis of all the follow-up
spectroscopy presented here. We like to thank the referee and A. Hatzes for
very useful comments.
The UVES spectra used in this analysis were obtained as part of an ESO
Service Mode run, proposal 269.C-5034. We acknowledge the use of the
\texttt{nightfall} program for the light-curve synthesis of eclipsing
binaries (http://www.lsw.uni-heidelberg.de/$\sim$rwichman/Nightfall.html),
written by Rainer Wichmann.  This research was supported in part by the DLR
under grant 50\,OR\,0201 (T\"ubingen) and in part by NSF grants AST-9720704
and AST-0086246, NASA grants NAG5-8425, NAG5-9222, as well as NASA/JPL
grant 961582 to the University of Georgia.  This work was supported in part
by the P\^ole Scientifique de Mod\'elisation Num\'erique at ENS-Lyon.  Some
of the calculations presented in this paper were performed on the IBM
pSeries 690 of the Norddeutscher Verbund f\"ur Hoch- und
H\"ochstleistungsrechnen (HLRN), on the IBM SP ``seaborg'' of the NERSC,
with support from the DoE, and on the IBM SP ``Blue Horizon'' of the San
Diego Supercomputer Center (SDSC), with support from the National Science
Foundation.  We thank all these institutions for a generous allocation of
computer time.

\end{acknowledgements}

\end{document}